\documentclass{PoS}

\title{
NLO QCD corrections for DVCS and TCS} 
\ShortTitle{NLO corrections for DVCS and TCS }

\author{H. Moutarde\\
      Irfu-SPhN, CEA, Saclay, France \\
}

\author{B. Pire\\
       CPHT, {\'E}cole Polytechnique, CNRS, 91128 Palaiseau, France \\
}

\author{F. Sabati{\'e}\\
       Irfu-SPhN, CEA, Saclay, France \\
}

\author{\speaker{L. Szymanowski}%\thanks{A footnote may follow.}
\\
       National Centre for Nuclear Research (NCBJ), Warsaw, Poland \\
        E-mail: \email{lechszym@fuw.edu.pl}}

\author{J. Wagner\\
      National Centre for Nuclear Research (NCBJ), Warsaw, Poland \\
}

\abstract{The inclusion of QCD corrections to the Born amplitude of deeply virtual Compton scattering in both spacelike (DVCS) and timelike (TCS) regimes modifies  the extraction process of generalized parton distributions (GPDs) from observables. In particular, gluon contributions are by no means negligible even in the medium energy range accessible at JLab12. We emphasize the complementarity of spacelike and timelike measurements and raise the question of factorization scale dependence of the $O(\alpha_S)$ results.}

\FullConference{Photon 2013,\\
		20-24 May 2013\\
		Paris, France }

\begin{document}

%\section{Introduction}

\begin{figure}[h]
\begin{center}
%\vspace*{-1cm}
%\hspace*{1.5cm}
\includegraphics[width=.44\textwidth, angle=0]{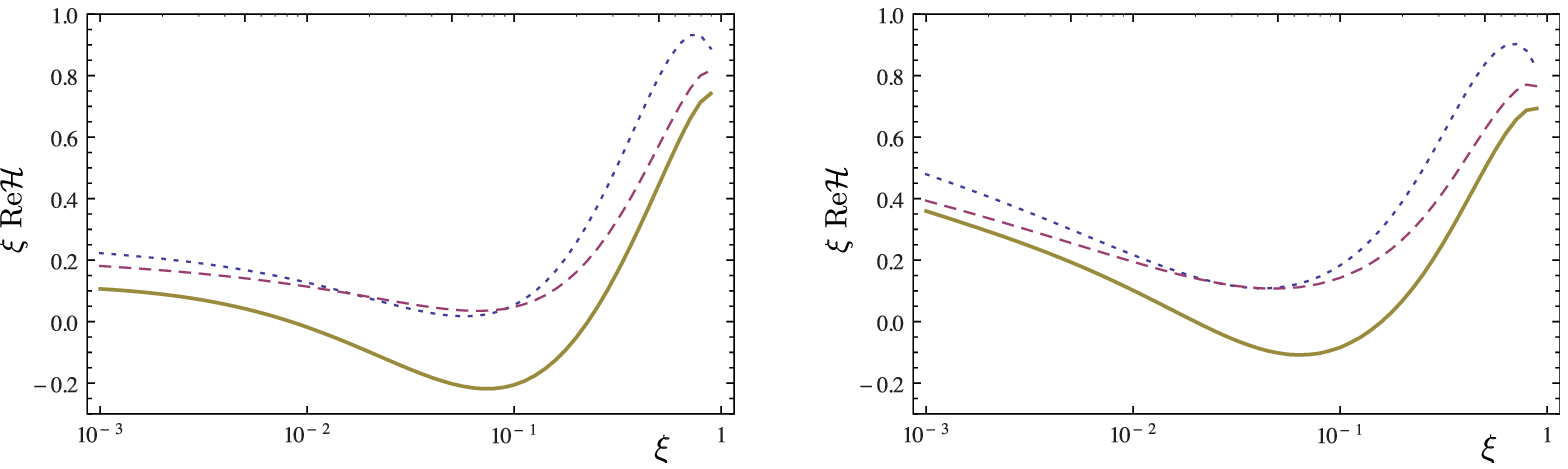}
\hspace*{0.4cm}
\includegraphics[width=.44\textwidth, angle=0]{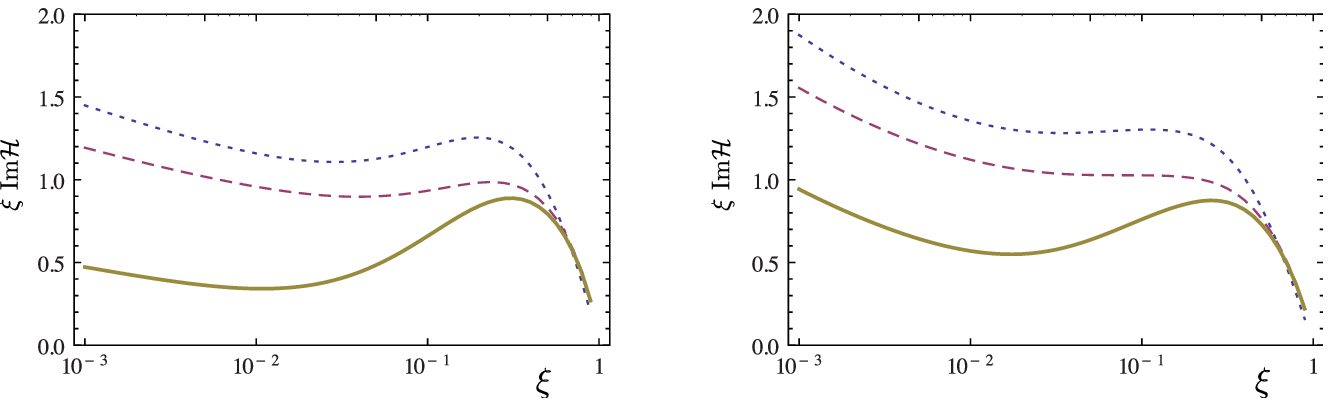}
\hspace*{.5cm}
%\includegraphics[width=.12\textwidth, angle=270]{fig5aImTCSHgk.eps}
%\hspace*{0.4cm}
%\includegraphics[width=.12\textwidth, angle=270]{fig5bImTCSHmstw.eps}
\caption{The real  (two left panels) and imaginary  (two right panels) parts of the spacelike DVCS Compton Form Factor ${\cal H}$ multiplied by $\xi$, 
as a function of $\xi$ in GK (first and third panels) and MSTW (second and fourth panels) double distribution models, for $\mu_F^2=Q^2=4$ GeV$^2$ and $t =-0.1$ GeV$^2$.
 In all plots, the LO result is shown as the dotted line, the full NLO result by the solid line and the NLO result without the gluonic contribution as the dashed line.
}
\end{center}
\end{figure}

\begin{figure}[h]
\begin{center}
%\vspace*{-1cm}
%\hspace*{1.5cm}
\includegraphics[width=.12\textwidth, angle=270]{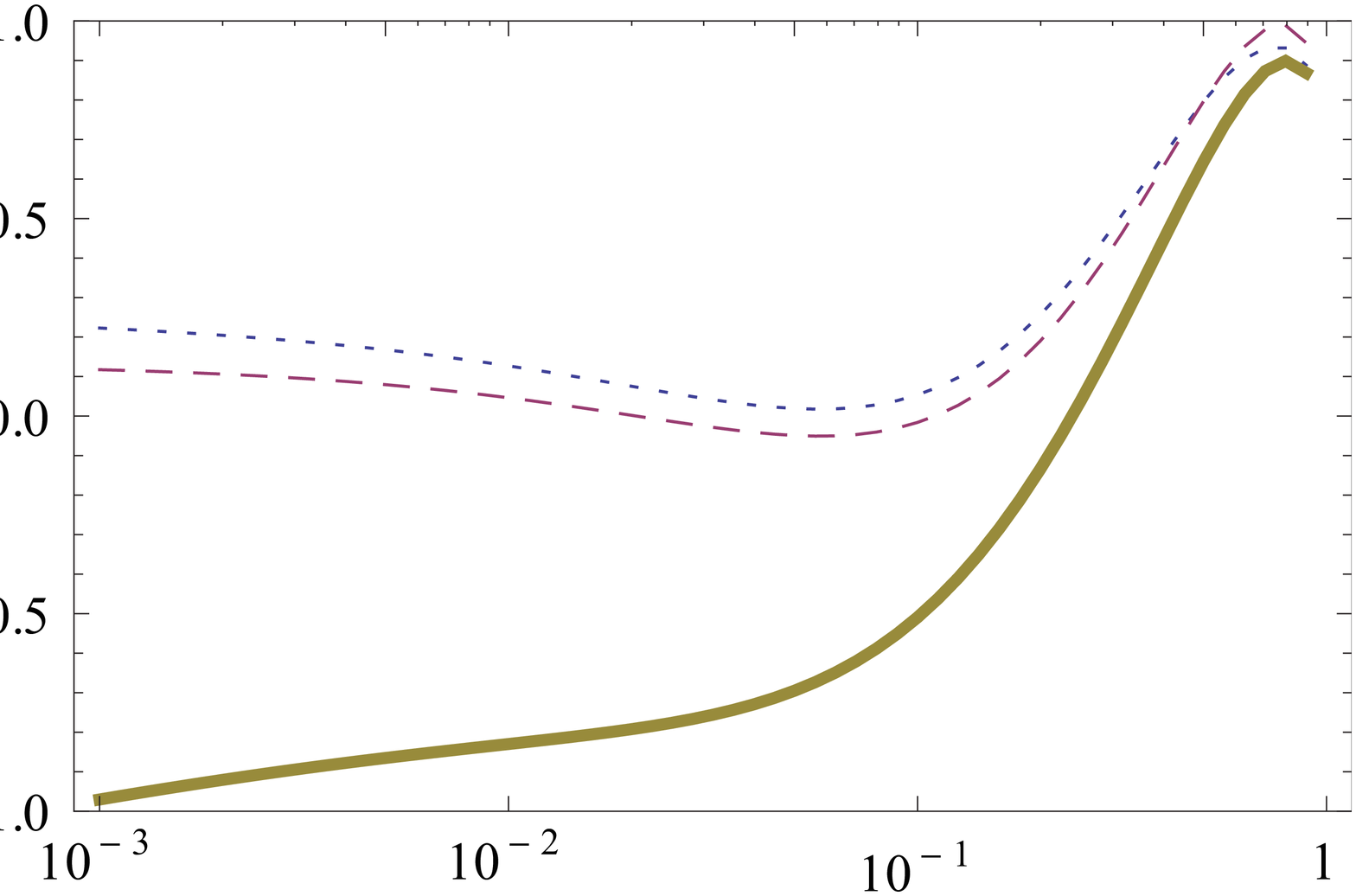}
\hspace*{0.4cm}
\includegraphics[width=.12\textwidth, angle=270]{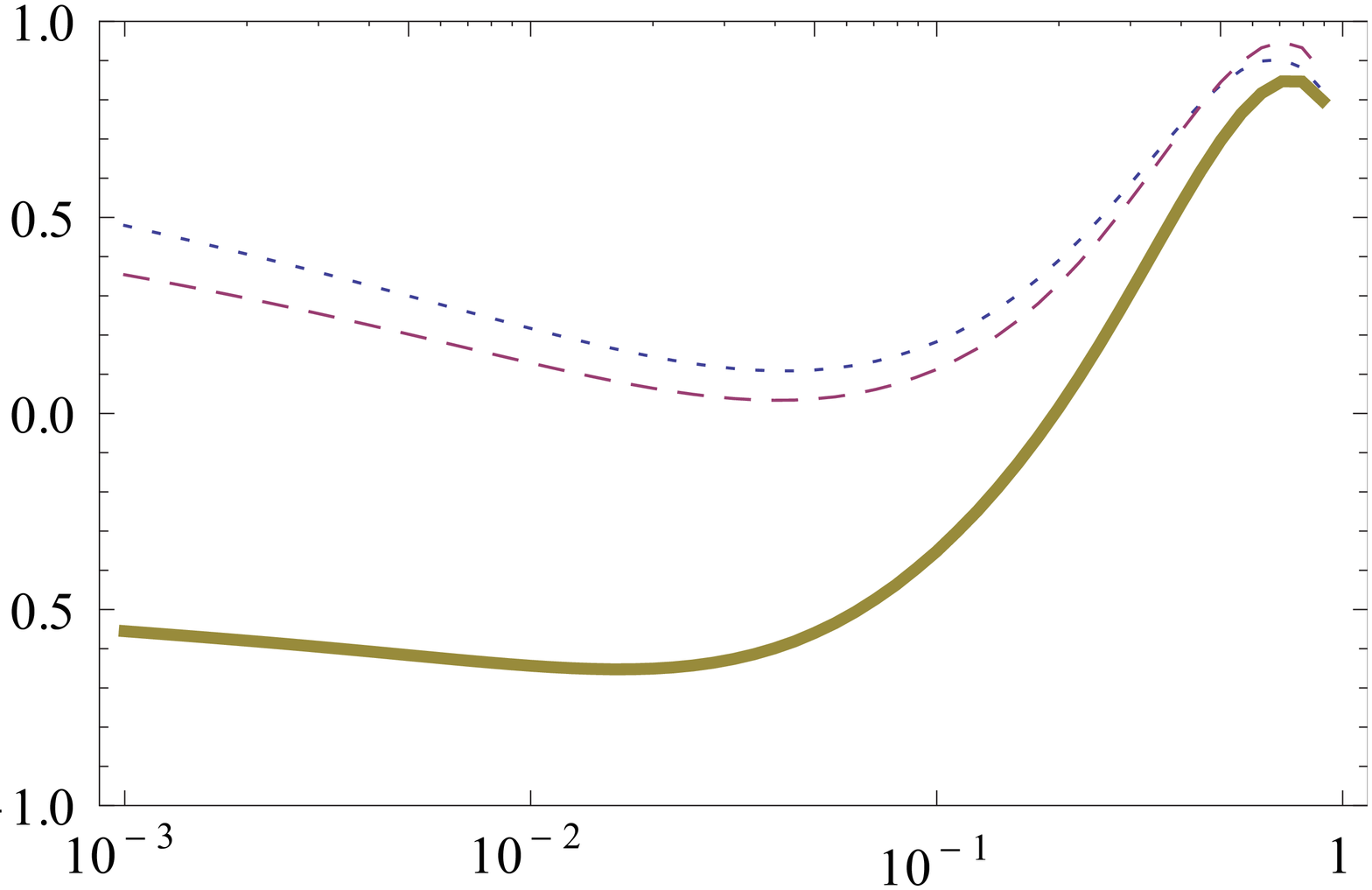}
\hspace*{.5cm}
\includegraphics[width=.12\textwidth, angle=270]{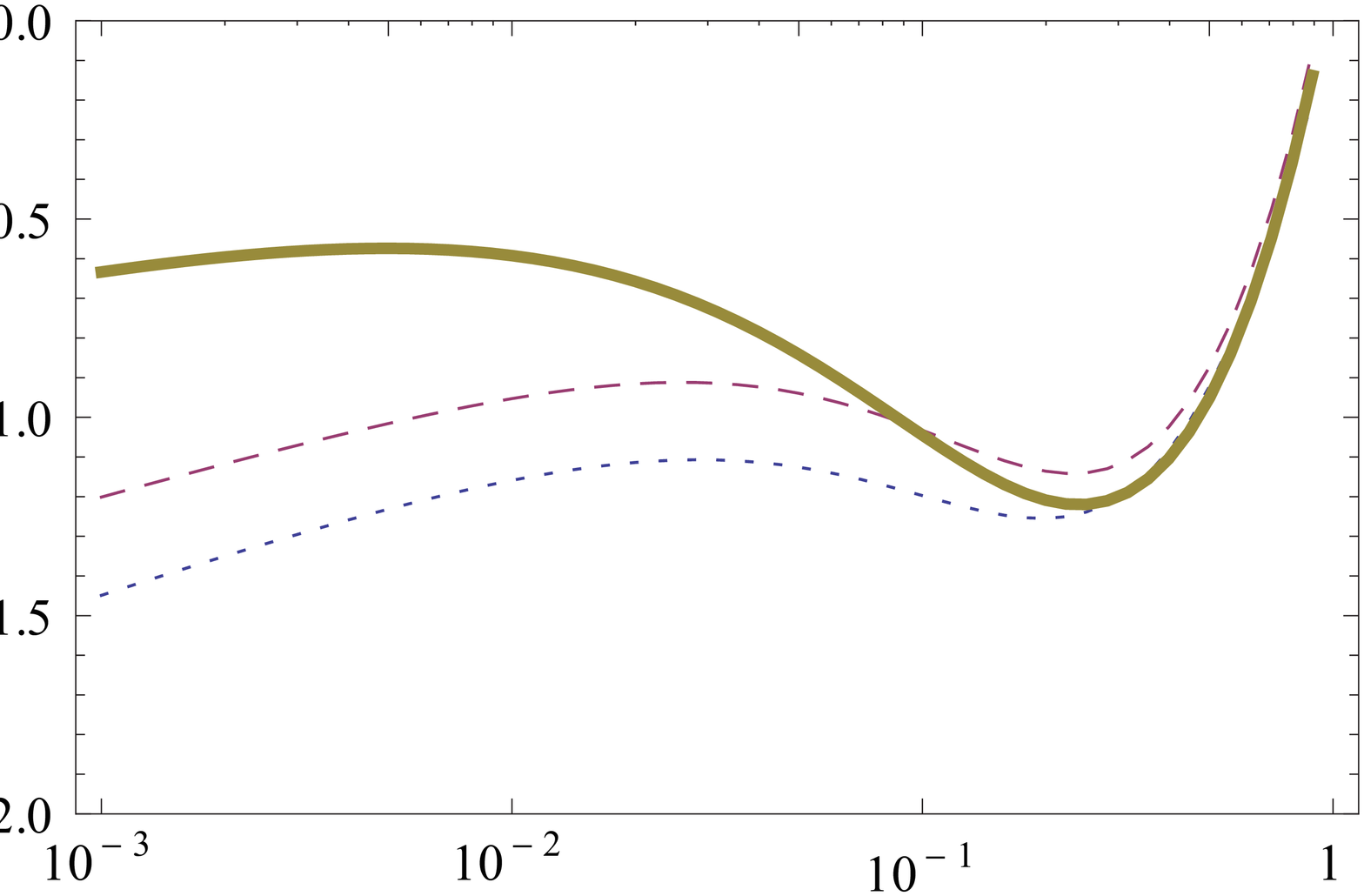}
\hspace*{0.4cm}
\includegraphics[width=.12\textwidth, angle=270]{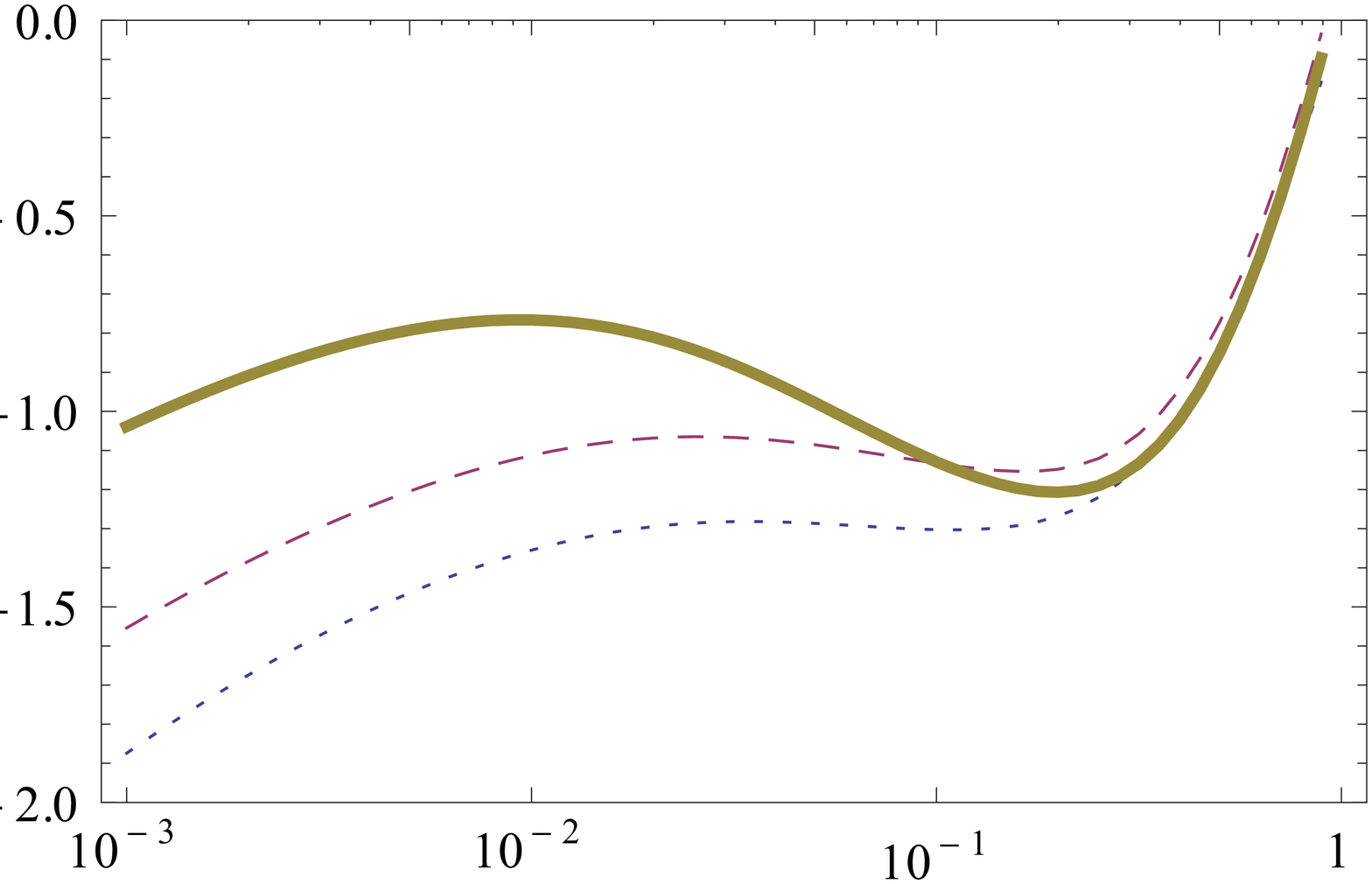}
\caption{The real  (two left panels) and imaginary  (two right panels) parts of the timelike TCS Compton Form Factor ${\cal H}$ multiplied by $\eta$, as a function of $\eta$ in GK (first and third panels) and MSTW (second and fourth panels) double distribution models, for $\mu_F^2=Q^2=4$ GeV$^2$ and $t =-0.1$ GeV$^2$.
}
\end{center}
\end{figure}

\begin{figure}[h]
%\vspace*{-1cm}
\hspace*{2cm}
 \includegraphics[width= 0.06\textwidth,angle=270]{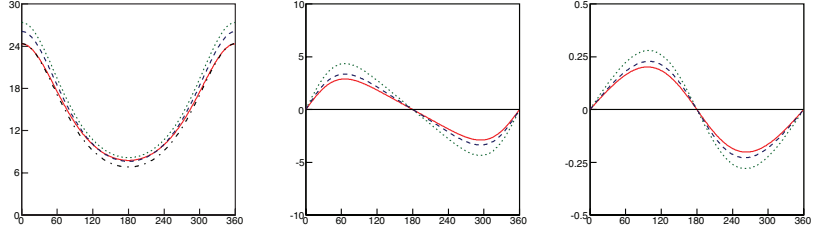}
%\includegraphics[width=.05\textwidth, angle=270]{fig9abcDVCSJLABgk.pdf}
%\hspace*{0.2cm}
%\includegraphics[width=.3\textwidth, angle=270]{fig2aImDVCSHgk.eps}
\vspace*{3cm}
\caption{From left to right, the total DVCS cross section in pb/GeV$^4$, the difference of cross sections for opposite lepton
helicities in pb/GeV$^4$, the corresponding asymmetry, all as a function of the usual $\phi$ angle (in Trento conventions \cite{TrentoC}) for
$E_e$ = 11 GeV; $\mu_F^2$ = $Q^2$ = 4 GeV$^2$ and t = - 0.2 GeV$^2$.  Curves correspond respectively to the  pure Bethe-Heitler contribution (dashed), the Bethe Heitler + interference at LO (dotted) and the Bethe-Heitler + interference at NLO (solid). }
\end{figure}

\begin{figure}[h]
%\vspace*{-1cm}
\hspace*{3cm}
\includegraphics[width=.4\textwidth, angle=0]{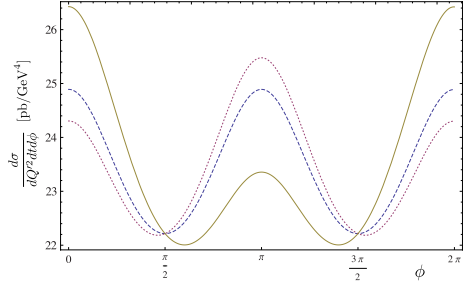}
%\hspace*{0.2cm}
%\includegraphics[width=.65\textwidth, angle=0]{xsec.eps}
%\vspace*{3cm}
\caption{
The $\phi$ dependence of the lepton pair photoproduction cross-section at $E_\gamma = 10$ GeV,  $Q^2 =  \mu ^2 = 4$~GeV$^2$,  and $t= -0.1$~GeV$^2$ integrated over $\theta \in (\pi/4,3\pi/4)$: pure Bethe-Heitler contribution (dashed), Bethe-Heitler plus interference contribution at LO (dotted) and NLO (solid).
}
\end{figure}

\begin{figure}[t]
%\vspace*{-1cm}
\hspace*{3cm}
\includegraphics[width=.4\textwidth, angle=0]{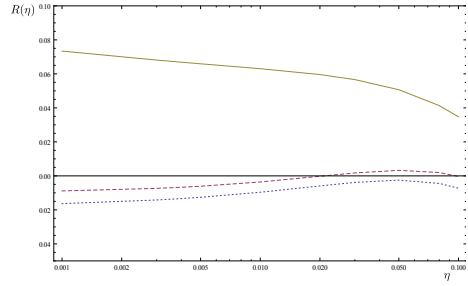}
%\hspace*{0.2cm}
%\includegraphics[width=.58\textwidth, angle=0]{xsec.eps}
%\vspace*{3cm}
\caption{The R ratio defined by Eq.  6 as a function of $\eta$, for $Q^2 = \mu_F^2 = 4$ GeV$^2$ and $t= -0.1$ GeV$^2$; 
the LO result is shown as the dotted line, the full NLO result by the solid line and the NLO result without the gluonic contribution as the dashed line.}
\end{figure}

Generalized parton distributions (GPDs) \cite{historyofDVCS, gpdrev} are a beautiful tool to access the 3-dimensional inner structure of hadrons \cite{Impact}. A necessary step to extract in a reliable way some information on  quark and gluon GPDs is to study  \cite{NLO}  $O(\alpha_s)$ QCD contributions to the amplitude of spacelike Deeply Virtual Compton Scattering (DVCS) :
\begin{equation}
\gamma^*(q_{in}) N(P) \to \gamma(q_{out}) N'(P'=P+\Delta) \,,~q_{in}^2 =-Q^2,~q_{out}^2 =0,~t=\Delta^2,~\xi =\frac{Q^2}{(P+P')\cdot(q_{in}+q_{out})} \nonumber \,,
\end{equation}
and of its crossed reaction, timelike Compton scattering (TCS) :
\begin{equation}
\gamma(q_{in}) N(P)\to \gamma^*(q_{out}) N'(P'=P+\Delta)\,,~q_{in}^2 =0,~q_{out}^2 =Q^2,~t=\Delta^2,~ \eta =\frac{Q^2}{(P+P')\cdot(q_{in}+q_{out})}  \nonumber \,.
\end{equation}
 After factorization, the DVCS (and similarly TCS)  amplitude is written in terms of  Compton form factors  (CFF) $\mathcal{H}$, $\mathcal{E}$ and $\widetilde {\mathcal{H}}$, $\widetilde {\mathcal{E}}$  as :
\begin{eqnarray}
\mathcal{A}^{\mu\nu}(\xi,t) =  \frac{- e^2}{(P+P')^+}\, \bar{u}(P^{\prime}) 
\Big[\,
  && g_T^{\mu\nu} \, \Big(
      {\mathcal{H}(\xi,t)} \, \gamma^+ +
      {\mathcal{E}(\xi,t)} \, \frac{i \sigma^{+\rho}\Delta_{\rho}}{2 M}
   \Big) 
  \nonumber\\ &
   +&i\epsilon_T^{\mu\nu}\, \Big(
    {\widetilde{\mathcal{H}}(\xi,t)} \, \gamma^+\gamma_5 +
      {\widetilde{\mathcal{E}}(\xi,t)} \, \frac{\Delta^{+}\gamma_5}{2 M}
    \Big)
\,\Big]\, u(P) \, ,
\label{eq:amplCFF}
\end{eqnarray}
with the CFFs defined, for instance in the cases of $\mathcal{H}(\xi,t)$ and $\widetilde {\mathcal{H}}(\xi,t)$,  as :
\begin{eqnarray}
\mathcal{H}(\xi,t) &=& + \int_{-1}^1 dx \,
\left(\sum_q T^q(x,\xi)H^q(x,\xi,t)
 + T^g(x,\xi)H^g(x,\xi,t)\right) \; ,
 \nonumber \\
\widetilde {\mathcal{H}}(\xi,t) &=& - \int_{-1}^1 dx \,
\left(\sum_q \widetilde {T}^q(x,\xi)\widetilde {H}^q(x,\xi,t) 
+\widetilde {T}^g(x,\xi)\widetilde {H}^g(x,\xi,t)\right).
\label{eq:CFF}
\end{eqnarray}

%\section{Results}
To estimate Compton Form Factors (CFF), we use  the  NLO calculations of the coefficient functions which have been calculated in the DVCS case in the early days of GPD studies and more recently for  the TCS case \cite{NLO}, the two results being simply related thanks to the analyticity (in $Q^2$) properties of the amplitude \cite{MPSW}:
%%%%%%%%%%%%%%%%%%%%%%%%%%%%%%%%%%%%%%%%%%%%%%%%%%%%%%%%%%%%%%%%%%%%%%%%%%%%%%%%
\begin{eqnarray}
^{TCS}T(x,\eta) = \pm \left(^{DVCS}T(x,\xi=\eta) +  i \pi C_{coll}(x,\xi = \eta)\right)^* \,,
\label{eq:TCSvsDVCS}
\end{eqnarray}
where the $+$~$(-)$ sign corresponds to the vector (axial) case.

\begin{figure}[t]
\begin{center}
 \includegraphics[width= 0.4\textwidth]{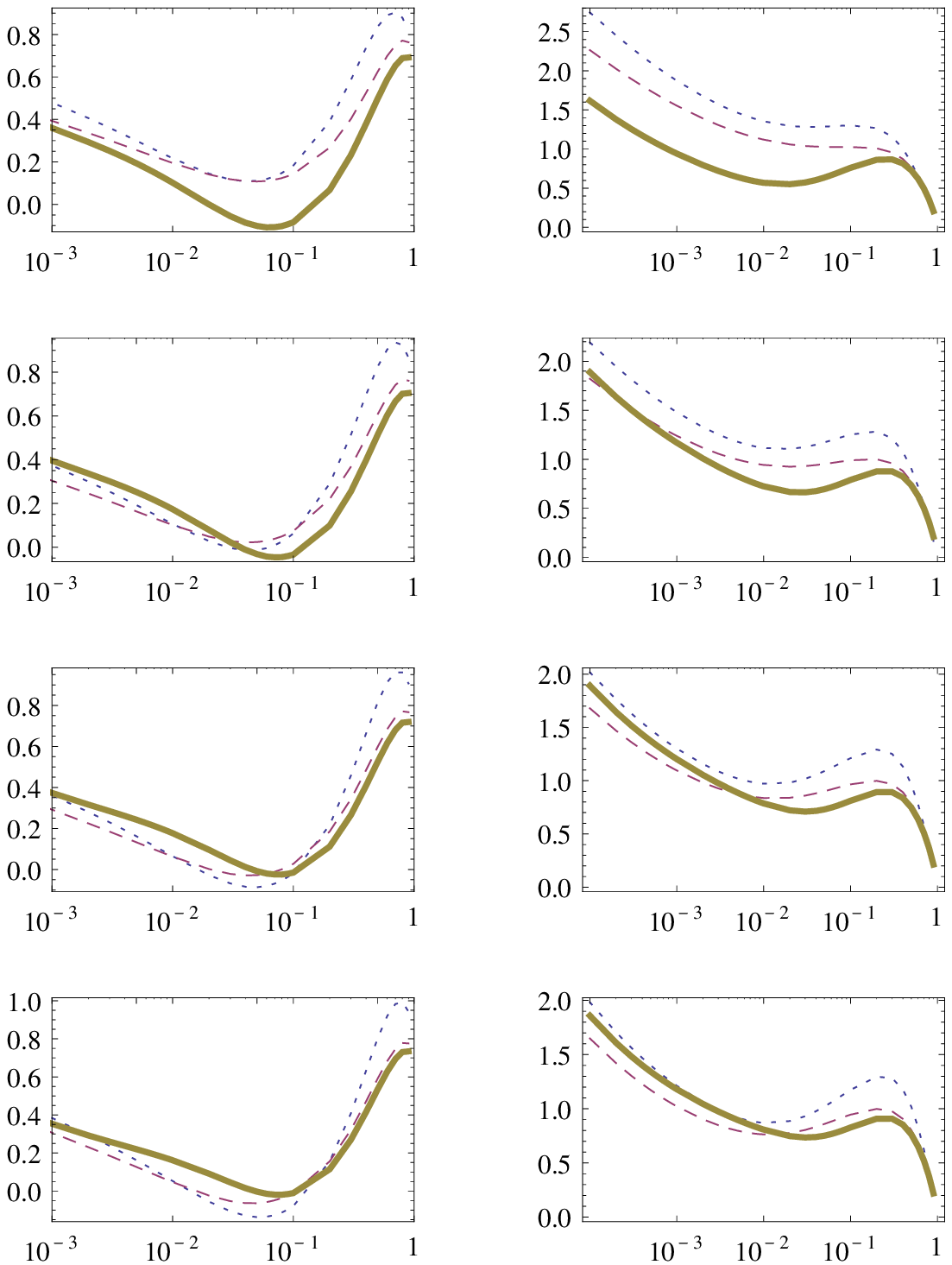} 
  \includegraphics[width= 0.4\textwidth]{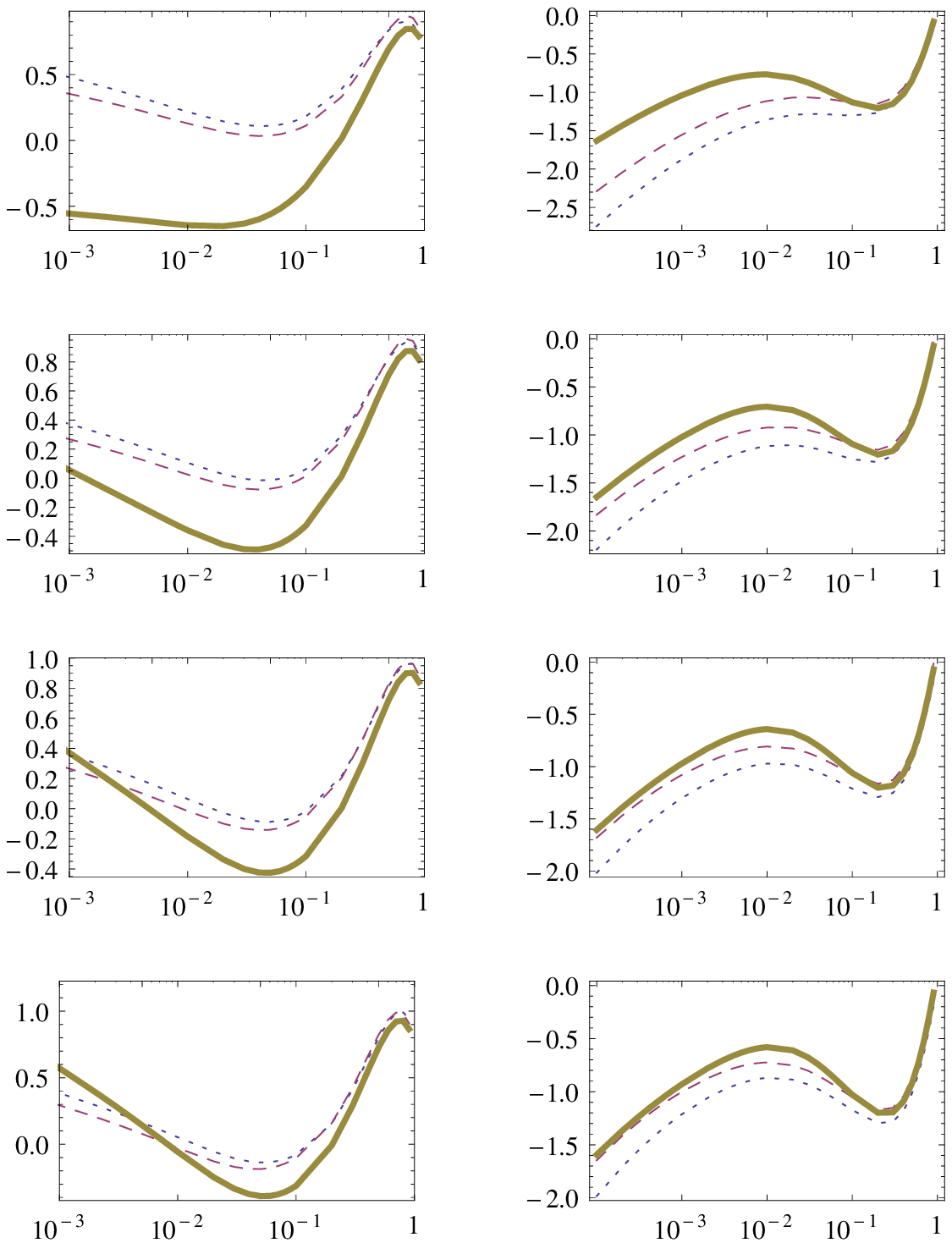}    
\caption{ The real (first and third columns) and imaginary (second and 
fourth columns) parts of the spacelike (first and 
second columns) Compton Form Factor $\xi \, \cal H$ and timelike (third and fourth columns) 
Compton Form Factor $\eta \, \cal H$ , for
 $\mu_F^2 = Q^2, Q^2/2, Q^2/3,  Q^2/4$, from top to bottom, 
and for $Q^2=4$ GeV$^2$, $t= -0.1$ GeV$^2$ and $\alpha_s=0.3$.}
%\label{Fig:crosssection2}
\end{center}
\end{figure}

Our estimates are based on two GPD models based on Double Distributions (DDs), as discussed in detail in Ref.  \cite{Moutarde:2013qs} : the Goloskokov-Kroll (GK) model  and a model (MSTW) based on the MSTW08 PDF parametrization. Our conclusions do not depend strongly on the GPD model used. 

We get the results shown in Fig. 1 and Fig. 2 for the real and imaginary parts of the spacelike and timelike dominant CFF $\mathcal{H}(\xi,t) $ and $\mathcal{H}(\eta,t) $, when choosing the factorization scale at the {\em natural} value $\mu_F^2=Q^2$.  Comparing dashed and solid lines  leads to the surprising observation that gluonic contributions are so important that they even change the sign of the real part of the CFF, and are dominant for almost all values of the skewness parameter. A milder conclusion arises for  the imaginary part of the CFF where the gluonic contribution remains sizeable for values of the skewness parameter up to $0.3$. 

Because of the competing Bethe Heitler mechanism which often dominates, the importance of NLO QCD corrections to observables depend on their sensitivity to the DVCS or TCS amplitudes. This is demonstrated in Fig. 3 in the DVCS case and in Fig. 4 and 5 for the TCS case. Note in particular the strong dependence of the ratio $R(\eta)$ defined \cite{BDP} as :

\begin{equation}
R(\eta)=\frac{2\int\limits_0^{2\pi}\;d\phi\;cos \phi\;\frac{d\sigma}{dQ^2\,dt\,d\phi}}{\int\limits_0^{2\pi}\;d\phi\frac{d\sigma}{dQ^2\,dt\,d\phi}},
\label{eq:Rratio}
\end{equation}
which is linear in the real part of the timelike CFF.

%\section{Conclusion}
The fact that both spacelike and timelike Compton form factors receive sizable  NLO contributions may worry the reader; indeed one usually tries to resum  large radiative corrections to stabilize a perturbative expansion. Although we  explored somewhat this possibility \cite{APSW}, we would like to prevent the critical reader   from drawing a hasty conclusion on the  convergence rate of the perturbative QCD expansion of the amplitude based on our NLO results. Indeed, most of the NLO correction comes from the gluonic term, which does not exist at LO. The large NLO contribution is therefore more a signature of the large size of the gluonic GPD than of the slow rate of the expansion. The real rate of the QCD expansion cannot be accessed before the NNLO contributions are computed. Our only measure of the validity of the QCD expansion is the smallness of the NLO quark contribution to the amplitude, as exemplified by the proximity of the dotted and dashed lines on Fig. 1 and 2.

Let us now turn to the factorization scale dependence of our results. There is no proven recipe to optimize the choice of the factorization scale in any QCD process. The question has been raised in several studies of inclusive and exclusive reactions but no definite strategy has yet emerged. In order to pave the way, we show on Fig. 6 the spacelike and timelike Compton form factor with the GK model, letting $\mu_F^2$ vary between $Q^2$ and $Q^2/4$.

\vskip.3in
\noindent
{\bf Acknowledgments}

\vskip.1in
\noindent
This work is partly supported by the Polish Grants NCN
No DEC-2011/01/D/ST2/02069, by the Joint Research Activity "Study of Strongly Interacting Matter"
(acronym HadronPhysics3, Grant Agreement n.283286) under the Seventh Framework Programme of the
European Community, by the GDR 3034 PH-QCD, and the ANR-12-MONU-0008-01, and by the
COPIN-IN2P3 Agreement.

\end{document}